\documentclass[twoside]{article}  %use this for latex2e
\usepackage{amsfonts}
\usepackage{amsbsy}
\usepackage{amsmath}
\usepackage{algorithmic}
\usepackage{algorithm}
\usepackage[pdftex]{graphicx}
\usepackage{subfig}
\newtheorem{thm}{Theorem}[section]
\newtheorem{lem}[thm]{Lemma}
\newtheorem{cor}[thm]{Corollary}
\newtheorem{example}[thm]{Example}
\newtheorem{defn}[thm]{Definition}
\newcommand{\pf}{\paragraph{Proof}}
\newcommand{\pfend}{\par\vspace{2ex}\noindent}
\newcommand{\eind}{\hspace*{\fill}$\Box$\par\vspace{2ex}\noindent}
\newcommand{\ruimte}{\par\vspace{1ex}\noindent}

\newcommand{\sfour}{ \;\;\;\; }

\newcommand{\Lag}{{\cal L}}

\newcommand{\F}{\mathbb{F}}

\newcommand{\N}{\mathbb{N}}

\newcommand{\beq}{\begin{equation}}
\newcommand{\eeq}{\end{equation}}
\newcommand{\bmat}{\left[ \begin{array}}
\newcommand{\emat}{\end{array} \right]}
\newcommand{\twee}[2]{\left[ #1 \sfour #2 \right]}
\newcommand{\COL}{\mathrm{col }\;}

\newcommand{\DIAG}{\mathrm{diag }\;}
\newcommand{\DEG}{\mathrm{deg }\;}
\newcommand{\WDEG}{\mathrm{wdeg }\;}

\newcommand{\RANK}{\mathrm{rank }\;}

\newcommand{\SPAN}{\;\mathrm{span }\;}

\newcommand{\ybold}{\mathbf{y}}
\newcommand{\rbold}{\mathbf{r}}
\DeclareMathOperator{\e}{e}

\DeclareMathOperator{\lt}{lt}
\DeclareMathOperator{\lc}{lc}
\DeclareMathOperator{\wdeg}{wdeg}

\DeclareMathOperator{\lm}{lm}
\DeclareMathOperator{\lpos}{lpos}

\title{\LARGE \bf
A parametric approach to list decoding of Reed-Solomon codes using interpolation}

\author{Mortuza Ali and Margreta Kuijper\footnote{M.\ Ali and M.\ Kuijper are with the Department of Electrical and Electronic Engineering, University of Melbourne, VIC 3010, Australia {\tt\small mortuzaa@unimelb.edu.au; mkuijper@unimelb.edu.au}}%
\thanks{This work was supported by the Australian Research Council(ARC).}}

\begin{document}
 
\maketitle
%%%%%%%%%%%%%%%%%%%%%%%%%%%%%%%%%%%%%%%%%%%%%%%%%%%%%%%%%%%%%%%%%%%%%%%%%%%%%%%%
\begin{abstract}
In this paper we present a minimal list decoding algorithm for Reed-Solomon (RS) codes. Minimal list decoding for a code $C$ refers to list decoding with radius $L$, where $L$ is the minimum of the distances between the received word $\mathbf{r}$ and any codeword in $C$. We consider the problem of determining the value of $L$ as well as determining all the codewords at distance $L$. Our approach involves a parametrization of interpolating polynomials of a minimal Gr\"obner basis $G$. We present two efficient ways to compute $G$. We also show that so-called re-encoding can be used to further reduce the complexity. We then demonstrate how our parametric approach can be solved by a computationally feasible rational curve fitting solution from a recent paper by Wu. Besides, we present an algorithm to compute the minimum multiplicity as well as the optimal values of the parameters associated with this multiplicity which results in overall savings in both memory and computation. 
\end{abstract}

\section{Introduction}
Reed-Solomon (RS) codes are important linear block codes that are of significant theoretical and practical interest. A $(n,k)$ RS code $C$, defined over a finite field $\F$, is a $k$ dimensional subspace of the $n$ dimensional space $\F^n$. For a message polynomial $m(x) = m_0 + m_1x+\cdots+m_{k-1}x^{k-1}$, the encoding operation is to evaluate $m(x)$ at $x_1, x_2, \ldots, x_n$, where the $x_i$'s are $n$ distinct elements of $\F$. The rich algebraic properties and geometric structures of RS codes lead to the invention of a number of efficient decoding algorithms such as Sugiyama algorithm~\cite{sugiyamaKHN75}, Berlekamp-Massey (BM) algorithm~\cite{berl68,massey69}, and Welch-Berlekamp (WB) algorithm~\cite{welchWB86}. These classical decoding algorithms guarantee correct decoding as long as the number of errors is upper bounded by $t = \lfloor (d-1)/2 \rfloor$, where $d = n-k+1$ is the minimum distance of the code.
\ruimte
In classical decoding, the error correcting radius of $t = \lfloor (d-1)/2 \rfloor$ originates from the requirement of unique decoding since for $t > \lfloor (d-1)/2 \rfloor$ multiple codewords within distance $t$ from the received word $\mathbf{r}$ may exist. One way to circumvent this limitation is to increase the decoding radius beyond $\lfloor (d-1)/2 \rfloor$ and allow the decoder to output a list of codewords rather than one single codeword. However, such list decoding is only feasible if there are few codewords in the list. In~\cite{guruthesis} Guruswami showed that for a code of relative distance $\delta = d/n$, any Hamming sphere of radius $\leq n(1 - \sqrt{1-\delta})$ around a received word $\mathbf{r}$ contains only a polynomial number of codewords. Therefore, a $(n,k)$ RS code with $d=n-k+1$ can be list decoded up to the error correcting radius of $n - \sqrt{n(k-1)}$ which Guruswami named as the Johnson bound.
\ruimte

A list decoding algorithm was first discovered for low rate RS codes by Sudan~\cite{sudan97a} and later improved and extended for all rates by Guruswami and Sudan~\cite{gursud99}. The Guruswami-Sudan algorithm can correct errors up to the Johnson bound $n - \sqrt{n(k-1)}$. Given a received word $\mathbf{r}$, the essential idea of the algorithm is to find all the polynomials $m$ of degree less than $k$ such that $m(x_i) \neq r_i$ for at most $t$ values of $i\in \{1, 2, \ldots, n \}$. The Guruswami-Sudan algorithm finds these polynomials in two steps: the interpolation step and the factorization step. In the interpolation step, it computes a bivariate polynomial $Q(x, r)$ that passes through all the points $(x_1, r_1), (x_2, r_2), \ldots, (x_n, r_n)$ with a prescribed multiplicity $s$ satisfying a certain weighted degree constraint (see~\cite{gursud99} for the definition of weighted degree). Then the bivariate polynomial $Q(x, r)$ is factorized to find all the factors of the form $r-m(x)$, where $m$ is a polynomial of degree less than $k$. Now a polynomial $m$ is a valid message polynomial if it is of degree less than $k$ and $m(x_i) \neq r_i$ for at most $t$ values of $i\in \{1, 2, \ldots, n \}$. The construction of $Q(x, r)$ with the prescribed multiplicity and weighted degree constraint ensures that for all valid message polynomials $m$, $r-m(x)$ appears as a factor of $Q(x, r)$. Even though the algorithm may produce implausible polynomials, the total number of polynomials $L$ in the list will satisfy the bound $L < (s+0.5)\sqrt{n/(k-1)}$, see~\cite{McEliece03}.

The most computationally intensive operation in the Guruswami-Sudan algorithm is the construction of the bivariate polynomial $Q(x, r)$. Computation of $Q(x, r)$ involves solving a system of $O(ns^2)$ homogeneous equations which using Gaussian elimination can be done in time cubic in the number of equations~\cite{trifonov07}. Clearly the algorithmic complexity of the interpolation step is dominated by the multiplicity $s$. Recently Wu~\cite{wu08} transformed the interpolation problem to a `rational interpolation problem' which involves smaller multiplicity. Given the received word $\mathbf{r}$, Wu's algorithm first computes the syndrome $\mathbf{s}$ of $\mathbf{r}$ followed by the computation of the error locator polynomial $\Lambda$ and error correction polynomial $B$ using the Berlekamp-Massey algorithm. Wu demonstrated that all valid error locator polynomials can be expressed as a parametrization of $\Lambda$ and $B$. More specifically, given a list decoding radius $t$, Wu's algorithm aims at finding all polynomials $\lambda$ and $\beta$ such that $\Lambda^\prime = \lambda \Lambda + \beta B$ has at most $t$ distinct roots. Wu showed that similar to the Guruswami-Sudan approach, this problem can be reduced to a curve fitting problem but with significantly smaller multiplicity.

It may be observed that the set of all $Q(x,r) \in \F[x, r]$ passing through the points $(x_i, r_i)$, for $i=1, 2, \ldots, n$, with multiplicity $s$ is an ideal $I_s$. From this observation several authors including Alekhnovich~\cite{alekhnovich05}, Nielsen and H\o holdt~\cite{nielsenH00}, Kuijper and Polderman~\cite{kuijppol_it}, O'Keeffe and Fitzpatrick~\cite{keeffeP07}, and Lee and O'Sullivan~\cite{leeS08}, formulated the interpolation step of the list decoding algorithm as the problem of finding the minimal weight polynomial from the ideal $I_s$. Clearly the minimal weight polynomial will appear as the minimal polynomial in a minimal Gr\"obner basis of $I_s$ computed with respect to the corresponding weighted term order. 
%% MA replaces this: However, the computation of a minimal Gr\"obner basis of an ideal is computationally complex. To overcome the computational difficulty, Lee and O'Sullivan employed the theory of Gr\"obner bases of modules. 
Lee and O'Sullivan also showed that the minimal polynomial in the ideal $I_s$ can be computed more efficiently from a minimal Gr\"obner basis of a submodule of $\F[x]^q$ for a sufficiently large $q$\footnote{Here the integer $q$ is not related to the size of the field.}.
Let $\F[x, r]_q = \{ f \in \F[x, r]\,|\,r\textnormal{-deg}(f) < q \}$. Then $\F[x, r]_q$ can be viewed as a free module over $\F[x]^q$ with a free basis $1, r, \ldots, r^{q-1}$. Then the essential observation of Lee and O'Sullivan is that the minimal polynomial of $I_s$ can be constructed from the minimal Gr\"obner basis of a submodule of $\F[x]^q$ along with the free basis $1, r, \ldots, r^{q-1}$, for large enough $q$.

%%Now let $I_{s, q} = I_s \cap F[x, r]_q$.
%%$I_{s, q}$ is a submodule of $F[x,r]_q$ and for large enough $q$ the minimal polynomial of $I_{s, q}$ is also the minimal polynomial of $I_s$.

In this paper we employ the theory of minimal Gr\"obner bases to perform minimal list decoding. Given the received word $\mathbf{r}$, let $L$ denote the value of $d_{H}(\mathbf{r}, C)$ where \[
d_{H}(\mathbf{r}, C) := \min_{\mathbf{c} \in C} \{ d_H(\mathbf{r}, \mathbf{c})\} .
\]
Our main objective is to determine the value of $L$ as well as all codewords $\mathbf{c}$ which are at a distance $L$ from the received word $\mathbf{r}$. Clearly, if $L$ is larger than the classical error correcting radius $\lfloor (d-1)/2 \rfloor$, the task is a list decoding operation. Our algorithm, unlike the Lee and O'Sullivan approach, starts with computing a minimal Gr\"obner basis $G$ of a submodule of $\F[x]^2$, rather than $\F[x]^q$ . We then demonstrate that all valid message polynomials can be extracted from a parametrization in terms of the elements of $G$. For computational feasibility, we show that this parametric approach, like Wu's algorithm, can be translated into a `rational interpolation problem'. However, our approach has at least three features that distinguishes it from Wu's algorithm. Firstly, our parametric formulation of the problem of list decoding, without the detour of syndrome computation, is simpler than Wu's formulation. Secondly, while Wu's algorithm, for each valid $\Lambda^\prime$, resorts to Forney's formula to compute the error values, our algorithm immediately leads to a valid message polynomial. Finally, we provide an algorithm to compute the minimum multiplicity along with the optimal values of the associated parameters to be used in the rational interpolation step. Use of these optimal parameters in the rational interpolation step results in savings of both memory and computation as compared to Wu's algorithm.

The organization of the rest of the paper is as follows. In Section~\ref{sec:preli}, we briefly review the relevant theory of Gr\"obner bases. In Section~\ref{sec:theory-alg}, we develop the theory and present the main algorithm along with two ways to compute the minimal Gr\"obner basis. In this section we also explain how so-called re-encoding can be applied to the proposed approach. In Section~\ref{sec:rational}, we translate the parametric approach into a `rational interpolation problem' and present an efficient algorithm for the computation of the minimum value of the multiplicity and other parameters to be used in the rational interpolation step. We demonstrate that the use of these optimal values of the parameters results in less memory requirement as well as less computational requirement as compared to Wu's approach. Finally we conclude the paper in Section~\ref{sec:con}.

%\begin{itemize}
%\item Sudan approach: multiplicity needs to be chosen high enough.
%\item in this paper maximum likelihood decoding. Known to be NP-hard.
%\item method based on interpolation rather than syndromes, going directly for message polynomial rather than error locator polynomial
%\item approach incorporates classical interpolation-based methods such as Welch-Berlekamp and Euclidean algorithm.
%\item In section .. we present our main approach which is conceptually simple. For the list decoding no decisions on multiplicity are needed. Given a received word $\mathbf{r}$, this method outputs a lower bound on the number of errors that can be decoded and then proceeds to produce a list of all codewords at minimum Hamming distance to $\rbold$. We present two versions, one based on the Euclidean algorithm and one based on an iterative algorithm. The method does not require a correct choice of parameter (multiplicity) as in the Sudan/Guruswami method. Depending on the received word, the actual list decoding may not be computationally feasible. For this reason we present a more computationally feasible algorithm in section ... which draws upon some recent ideas from ...
%\item If re-encoding is used, the iterative option essentially uses the four polynomials that are output by a Welch-Berlekamp algorithm.
%\end{itemize}

\section{Preliminaries}
\label{sec:preli}
The theory of Gr\"obner bases for modules in $\F[x]^q$ is generally
recognized as a powerful conceptual and computational tool that plays
a role similar to Euclidean division for modules in $\F[x]$. More specifically, minimal Gr\"obner bases prove themselves as an effective tool for various types of interpolation problems. In recent papers~\cite{kuijperS09,kuijperS10} this effectiveness was ascribed to a powerful property of minimal Gr\"obner bases, explicitly identified as the `Predictable Leading Monomial Property'. The proofs in this paper make use of this property. Before recalling the PLM property let us first recall some terminology on Gr\"obner bases.
\ruimte
Let $\e_1, \dots, \e_q$ denote the unit vectors in $\F^q$. 
The elements $x^\alpha \e_i$ with $i \in \{1, \dots, l\}$ and 
$\alpha \in \N_0$ are called \textbf{monomials}.
Let $n_1, \ldots , n_q$ be nonnegative integers. In this paper we define the following two types of monomial orders:
\begin{itemize}
 \item[$\bullet$] The {\bf $\boldsymbol{(n_1,\cdots,n_q)}$-weighted term over position ({\tt top})} order, defined as 
$$ x^{\alpha} \e_i < x^{\beta} \e_j \;\; :\Leftrightarrow \;\; \alpha + n_i < \beta +n_j
\mbox{ or } ( \alpha + n_i =\beta +n_j \mbox{ and } i<j ). $$ 
 \item[$\bullet$] The {\bf $\boldsymbol{(n_1,\cdots,n_q)}$-weighted position over term ({\tt pot})} order, defined
 as 
$$ x^{\alpha} \e_i < x^{\beta} \e_j \;\; :\Leftrightarrow \;\; i<j \mbox{ or } (
 i=j \mbox{ and }  \alpha + n_i < \beta +n_j ) .$$  
\end{itemize}
Clearly, whatever order is chosen, every nonzero element $f \in \F[x]^q$ can be written uniquely as
$$ f=\sum_{i=1}^L c_i X_i, $$ where $L \in \N$, the $c_i$'s are nonzero
elements of $\F$ for $i=1 , \ldots , L$ and the polynomial vectors $X_1, \ldots
 , X_L$ are monomials, ordered as $X_1 > \dots > X_L$. Using the terminology of~\cite{AdamsLoustaunau} 
we define
\begin{enumerate}
\item[$\bullet$] $\lm(f):=X_1$ as the \textbf{leading monomial} of $f$
\item[$\bullet$] $\lt(f):=c_1X_1 $ as the \textbf{leading term} of $f$
\item[$\bullet$] $\lc(f):=c_1$ as the \textbf{leading coefficient} of $f$
\end{enumerate}
Writing $X_1=x^{\alpha_1} \e_{i_1}$, where $\alpha_1 \in \N_0$ and $i_1
\in \{1,\ldots , l \}$, we define
\begin{enumerate}
 \item[$\bullet$] $\lpos(f):=i_1$ as the \textbf{leading position} of $f$
 \item[$\bullet$] $\wdeg(f):=\alpha_1 + n_{i_1}$ as the \textbf{weighted degree} of $f$.
\end{enumerate}
Note that for zero weights $n_1=\cdots=n_q=0$ the above orders coincide with the reflected versions of the standard TOP order and POT order, respectively, as introduced in the textbook~\cite{AdamsLoustaunau}.
\ruimte
Also note that, unlike with TOP, the introduction of weights does not change the POT ordering of monomials. In this paper, weighted POT order is needed only because we need the associated notion of `weighted degree'.
\ruimte

We now recall some basic definitions and results on Gr\"obner bases, see~\cite{AdamsLoustaunau}. 
Below we denote the
submodule generated by a polynomial vector $f$ by $\langle f \rangle$.
\begin{defn}
Let $F$ be a subset of $\F [x]^q$. Then the submodule $\Lag(F)$, defined as
$$\Lag(F):= \langle \lt(f) \ | \ f \in F \rangle$$
is called the \textbf{leading term submodule} of $F$.
\end{defn}
%For example, for $q=2$, let $F=\{ \left[ x^2 \;\;\; x^3 \right]  \}$.
%Using $\TOP$ we obtain $\Lag(F) = \langle \left[ 0 \;\;\; x^3 \right]
%\rangle$, whereas using $\POT$ we get $\Lag(F) = \langle \left[ x^2
%  \;\;\; 0 \right] \rangle$. 
\begin{defn}\label{def_grob}
Let $M \subseteq \F[x]^q$ be a module and $G \subseteq M$. Then
$G$ is called a \textbf{Gr\"obner basis} of $M$ if $$ \Lag(G) = \Lag(M).$$
\end{defn}
In order to define a concept of minimality we have the following definition.
\begin{defn}\label{DefRed}{\rm (\cite[Def.\ 4.1.1]{AdamsLoustaunau}) }
 Let $0 \neq f \in \F[x]^q$ and let $F=\{f_1, \dots, f_s\}$ be a set of nonzero elements of $\F[x]^q$. Let $\alpha_{j_1}, \dots, \alpha_{j_m} \in \N_0$ and $\beta_{j_1}, \dots, \beta_{j_m}$ be nonzero elements of $\F$, where $1 \leq m \leq s$ and $1 \leq j_i \leq s$ for $i=1, \ldots, m$, such that
\begin{enumerate}
\item $\lm(f)=x^{\alpha_{j_i}}\lm(f_{j_i})$ for $i = 1,\ldots , m$ and
\item $\lt(f)=\beta_{j_1} x^{\alpha_{j_1}}\lt(f_{j_1}) + \dots + \beta_{j_m} x^{\alpha_{j_m}}\lt(f_{j_m})$.
\end{enumerate}
Define 
\[
h := f- (\beta_{j_1} x^{\alpha_{j_1}} f_{j_1} + \dots + \beta_{j_m} x^{\alpha_{j_m}} f_{j_m}) .
\]
Then we say that $f$ \textbf{reduces} to $h$ modulo $F$ in one step and we write 
$$ f \xrightarrow{F} h . $$
If $f$ cannot be reduced modulo $F$, we say that $f$ is
\textbf{minimal} with respect to $F$. 
\end{defn}
\begin{lem}\label{lemma_smaller}{\rm (\cite[Lemma 4.1.3]{AdamsLoustaunau}) }
Let $f$, $h$ and $F$ be as in the above definition. If $f
\xrightarrow{F} h$ then $h=0$ or $\lm(h)<\lm(f)$. 
\end{lem}
\begin{defn}{\rm (\cite{AdamsLoustaunau})}
 A Gr\"obner basis $G$ is called \textbf{minimal} if all its elements $g$ are minimal with respect to $G\backslash\{g\}$.
\end{defn}
It is well known~\cite[Exercise 4.1.9]{AdamsLoustaunau} that a minimal Gr\"obner basis
exists for any module in $\F[x]^q$ and that all leading positions of its elements are different.
In~\cite{kuijperS09,kuijperS10} another important property of a minimal Gr\"obner basis is identified; the theorem below merely formulates a well known result. 
\begin{thm}[~\cite{kuijperS09}]\label{thm_main_field}
Let $M$ be a submodule of $\F[x]^q$ with
minimal Gr\"obner basis $G=\{g_1 , \ldots , g_m \}$. Then for any $0 \neq f \in M$, written as 
\beq
f=a_1g_1 + \dots + a_mg_m ,\label{eq_lin_comb_f}
\eeq
where $a_1, \dots, a_m \in \F[x]$, we have
\beq
\lm(f)=\max_{1 \leq i \leq m; a_i \neq 0} ( \lm(a_i) \lm(g_i)) .\label{eq_plm}
\eeq
\end{thm}
The property outlined in the above theorem is called the \textbf{Predictable Leading Monomial (PLM) property}, as in~\cite{kuijperS09}. Note that this property involves not only degree information (as in the `predictable degree property' first introduced in~\cite{forney70}) but also leading position information. Most importantly, the above theorem holds irrespective of which monomial orders {\tt top} or {\tt pot} is chosen, for a proof see~\cite{kuijperS09}.
\ruimte
Clearly, in the above theorem $m=\RANK(M)$ and all minimal Gr\"obner bases of $M$ must have $\RANK(M)$ elements, no matter
 which monomial order is chosen. Furthermore, we have the following theorem.
\begin{thm}\label{thm_sum}
Let $n_1, \ldots , n_q$ be nonnegative integers and let $M$ be a module in $\F[x]^q$. Let $G=\{g_1 , \ldots , g_m \}$ be a minimal Gr\"obner basis of $M$ with 
respect to the $(n_1 ,\cdots , n_q)$-weighted {\tt top} order; denote $\ell_i := \wdeg 
g_i$ for $i=1, \ldots , m$. Let $\tilde G =\{\tilde g_1 , \ldots , \tilde g_m \}$ be a minimal Gr\"obner basis of $M$ with respect to the $(n_1 ,\cdots , n_q)$-
weighted {\tt pot} order; denote $\tilde \ell_i := \wdeg \tilde g_i$ for $i=1, \ldots ,m$. Then
\beq
\sum_{i=1}^m \ell_i = \sum_{i=1}^m \tilde \ell_i .\label{eq_ell}
\eeq
\end{thm}
\pf We first prove the theorem for the case $m=q$. It follows easily from the fact that both $G$ and $\tilde G$ are bases for $M$ (in a linear algebraic sense) that there exists a unimodular polynomial matrix $U \in \F[x]^{q\times q}$ such that 
\[
\COL\{g_1 , \ldots , g_q \} = U \COL\{\tilde g_1 , \ldots , \tilde g_q \} .
\]
Without restrictions we may assume that the leading positions within each Gr\"obner basis are strictly increasing. Clearly it follows from the above equation that also
\beq
V=UW ,\label{eq_vuw}
\eeq
where $V=\COL\{g_1 , \ldots , g_q \} \DIAG \{x^{n_1} , \cdots , x^{n_q}\}$ and $W=\COL\{\tilde g_1 , \ldots , \tilde g_q \} \DIAG \{x^{n_1} , \cdots , x^{n_q}\}$
Since $U$ is unimodular we must have deg det $V =$ deg det $W$. Clearly deg det $V =\sum_{i=1}^m \ell_i$ and deg det $W = \sum_{i=1}^m \tilde \ell_i$ from which~(\ref{eq_ell}) follows.
Next, we prove the general case $m \leq q$. For this, we note that it follows immediately from~(\ref{eq_vuw}) that the maximum degree of all minors of $V$ equals the maximum degree of all minors of $W$. On the other hand, the maximum degree of all minors of $V$ clearly equals $\sum_{i=1}^m \ell_i$ and similarly the maximum degree of all minors of $W$ equals $\sum_{i=1}^m \tilde \ell_i $. The theorem now follows.
\eind\pfend
We call the sum in~(\ref{eq_ell}) the {\bf $\boldsymbol{(n_1,\cdots,n_q)}$-weighted degree} of $M$, denoted by $\WDEG (M)$.
For zero weights $n_1=\cdots =n_q=0$ the above result expresses that the sum of 
the degrees of a (reflected) TOP minimal Gr\"obner basis of a module $M$ coincides with the 
sum of the degrees of a (reflected) POT minimal Gr\"obner basis of $M$. This result is merely a reformulation of the well known fact that the McMillan degree of a row reduced polynomial matrix equals the sum of its row degrees, see~\cite{forney75}.
\begin{cor}\label{cor_sum}
let $M$ be a module in $\F[x]^q$. Let $G=\{g_1 , \ldots , g_m \}$ be a Gr\"obner basis of $M$ whose $(n_1 ,\cdots , n_q)$-weighted {\tt top} degrees add up to $\WDEG (M)$. Then $G$ is a minimal Gr\"obner basis of $M$ with 
respect to the $(n_1 ,\cdots , n_q)$-weighted {\tt top} order.
\end{cor}
\pf
Suppose that $G$ is not minimal. Then there exists $g \in G$ that can be reduced modulo $G\backslash\{g\}$.
This implies that there exists a Gr\"obner basis of $M$ whose sum of weighted degrees is strictly less than $\WDEG (M)$, which contradicts the above theorem.
\eind
%\pfend
\section{Minimal list decoding through division}
\label{sec:theory-alg}
Let us now consider a $(n,k)$ RS code and a nonnegative integer $t$.
%define $\tau:= \lfloor (n-k)/2 \rfloor$ as its classical error bound.
The problem of `list decoding up to $t$ errors' is the following:
\ruimte
{\bf List Decoding Problem}: Given a received word $(r_1 , \cdots , r_n ) \in \F^n$, find all polynomials $m \in \F[x]$ of degree $<k$ such that 
\[
m(x_i ) = r_i \;\;\; \mbox{for at least $n-t$ values of }i \in \{ 1, \ldots , n \} .
\]
\subsection{Main approach}
We introduce the following two polynomials in $\F[x]$:
\beq
\Pi(x) = \prod_{i=1}^n(x-x_i) \label{eq_pi},
\eeq
and $\Lag$ as the Lagrange interpolating polynomial, i.e., the polynomial of least degree for which 
\beq
\Lag(x_i) = r_i \;\mbox{for all }i \in \{ 1, \ldots , n \} .\label{eq_Lagrange}
\eeq
\begin{defn}
Let $\mathbf{r} = ( r_1 , \cdots , r_n ) \in \F^n$. The {\bf interpolation module} $M (\mathbf{r} )$ is given by the module in $\F[x]^2$ that is spanned by the vectors $\tilde g_1 := \twee{\Pi(x)}{0}$ and $\tilde g_2 := \twee{\Lag(x)}{-1}$. 
\end{defn}
Note that $\{ \tilde g_1 ,\tilde g_2 \}$ is a minimal {\tt pot} Gr\"obner basis for $M (\mathbf{r} )$.
The above defined interpolation module is crucial to our approach. With $\tilde g_2$ we associate the bivariate polynomial $Q_2(x,y)=\Lag(x) - y$; clearly $Q_2(x_i , r_i)=0$ for all $i\in \{ 1, \ldots , n \}$. Similarly, with $\tilde g_1$ we associate the polynomial $Q_1(x,y)=\Pi(x)$; trivially $Q_1(x_i , r_i)=0$ for all $i\in \{ 1, \ldots , n \}$. Now consider an arbitrary bivariate polynomial $Q$ of the form $Q(x,y)=N(x) - D(x)y$ for which $Q(x_i , r_i)=0$ for all $i\in \{ 1, \ldots , n \}$. It can be shown, see~\cite{kuijppol_it}, that $\twee{N}{-D} \in M (\mathbf{r} )$. 
Recall that list decoding up to $t$ errors amounts to finding all polynomials $m \in \F[x]$ of degree $<k$ such that 
\[
m(x_i ) = r_i \;\;\; \mbox{for all $i \in \{ 1, \ldots , n \} $ except $i= j_1 , \ldots , j_L$ with $L \leq t$.}
\]
In our context this amounts to looking for an interpolating bivariate polynomial $Q$ of the form $Q(x,y)=D(x)m(x) - D(x)y$, where $D(x)=\prod_{i =1}^L (x- x_{j_i})$. Note that then indeed $Q(x_i , r_i)=0$ for {\em all} $i\in \{ 1, \ldots , n \}$.
Thus, to solve the above list decoding problem we are looking for particular vectors $\twee{N}{-D} \in M (\mathbf{r} )$ of weighted $(0,k-1)$-degree $\leq t+k-1$, that satisfy
\begin{enumerate}
\item $N$ is a multiple of $D$ and
\item $D$ has $L$ distinct zeros in $\F$, where $L$ denotes deg $D$.
\end{enumerate}
In this paper we are interested in finding the smallest value $L=d_H(\mathbf{r}, C)$ for which list decoding is possible as well as performing the associated list decoding. Thus we occupy ourselves with maximum likelihood list decoding. We have the following theorem. 
\begin{thm}\label{thm_req}
Let $\mathbf{r} = ( r_1 , \cdots , r_n ) \in \F^n$ be a received word and let $M (\mathbf{r} )$ be the corresponding interpolation module. Let $f=\twee{f^{(1)}}{f^{(2)}} \in \F [x]^2$ be a vector in $M (\mathbf{r} )$ of weighted $(0,k-1)$-degree $L$ that satisfies the following 3 requirements: 
\begin{enumerate}
\item $\lpos (f) =2$,
\item $f^{(1)}$ is a multiple of $f^{(2)}$ and
\item there is no vector in $M (\mathbf{r} )$ of weighted $(0,k-1)$-degree $< L$ that satisfies requirements 1) and 2).
\end{enumerate}
Then
\[
m:= - \frac{f^{(1)}}{f^{(2)}}
\]
is a message polynomial corresponding to a minimal error pattern of $L-k+1$ errors.
\end{thm}
\pf
From $\lpos (f) =2$ it follows immediately that deg $m < k$ and deg $f^{(2)} = L-k+1$. It remains to prove that $f^{(2)}$ has $L-k+1$ distinct zeros in $\F$. Since $f\in M (\mathbf{r} )$ there exist polynomials $\alpha$ and $\beta$ such that
\beq
f=\twee{\alpha}{\beta} \bmat{cc}\Pi & 0 \\ \Lag & -1 \emat . \label{eq_piL}
\eeq
Observe that $\alpha$ and $\beta$ do not have a common factor, otherwise the weighted degree of $f$ would not be minimal (requirement~3). From~(\ref{eq_piL}) it follows that $\alpha \Pi - f^{(2)} \Lag = f^{(1)}$ is a multiple of $f^{(2)}$ by requirement~2. As a result, $\alpha \Pi$ is a multiple of $f^{(2)}$. Since $\alpha$ and $\beta = -f^{(2)}$ have no common factor it follows that $\Pi$ must be a multiple of $f^{(2)}$, i.e., $f^{(2)}$ has $L-k+1$ distinct zeros in $\F$, which proves the theorem.
\eind
%\pfend
\begin{lem}\label{lem_para}
Let $\mathbf{r} = ( r_1 , \cdots , r_n ) \in \F^n$ be a received word and let $M (\mathbf{r} )$ be the corresponding interpolation module. Let $\{ g_1 , g_2 \}$ be a $(0,k-1 )$-weighted {\tt top} minimal Gr\"obner basis for  $M (\mathbf{r} )$ with $\lpos (g_2 ) = 2$. Denote $\ell_1 := \wdeg g_1$ and $\ell_2 := \wdeg g_2$. Let $t$ be a nonnegative integer. Then a parametrization of all vectors $f \in \F [x]^2$ with $\lpos (f) = 2$ and $\wdeg f = t+k-1$ (with respect to the $(0,k-1 )$-weighted {\tt top} order) is given by
\[
f = a g_1 + b g_2 ,
\]
where $a \in \F [x]$ with $\deg a \leq t+k-1-\ell_1$ and $b$ is a monic polynomial in $\F [x]$ of degree $t+k-1-\ell_2$. In particular, there exist no such vectors $f$ for $t < \ell_2 -k+1$.
\end{lem}
\pf
According to Theorem~\ref{thm_main_field}, $\{ g_1 , g_2 \}$ has the PLM property with respect to the $(0,k-1 )$-weighted {\tt top} order. The parametrization now follows immediately from this property.
\eind\pfend
%Note: no ambiguity as in Wu p.3616 regarding which polynomial has largest degree  STILLTOWRITE.
%\ruimte
Together, the above lemma and theorem give rise to the heuristic list decoding Algorithm~\ref{alg_main}.
\begin{algorithm}[!htb]
%\begin{algorithm}[!htb]
\caption{Minimal list decoding of $(n, k)$ RS code}
\label{alg_main}
\begin{algorithmic}
\STATE
\STATE {\bf Input}: Received word $\mathbf{r}= (r_1, \ldots, r_n)$
\STATE {\bf Output}: A list of polynomials $m$ of degree $< k$ such that $d_H(\mathbf{c}, \mathbf{r})$ is minimal, where $\mathbf{c}=(m(x_1), \ldots, m(x_n))$.
\STATE
\STATE 1. Compute the polynomials $\Pi$ and $\Lag$ given by~(\ref{eq_pi}) and~(\ref{eq_Lagrange}) ; define the interpolation module $M(\mathbf{r}) :=  \SPAN \{ \twee{\Pi}{0} , \twee{\Lag}{-1}\}$.
\STATE 2. Compute a minimal Gr\"obner basis $G = \{g_1 , g_2\}$ of $M(\mathbf{r}) $ with respect to the $(0,k-1 )$-weighted {\tt top} monomial order, with $\lpos (g_2) = 2$. Denote $\ell_1 := \wdeg g_1$ and $\ell_2 := \wdeg g_2$; set $j=0$. 
\STATE 3. Check requirement 2) of Theorem~\ref{thm_req} for $f =a g_1 + b g_2$, for all $a \in \F [x]$ with $\deg a \leq \ell_2 - \ell_1 +j$ and for all monic $b \in \F [x]$ with $\deg b = j$; write $f = \twee{f^{(1)}}{f^{(2)}}$.
\STATE 4. Whenever step 3) is successful, output all obtained quotient polynomials, i.e., polynomials $m$ of the form $m= -f^{(1)} / f^{(2)}$. In case step 3) is not successful increase $j$ by $1$ and repeat step 3).
\end{algorithmic}
\end{algorithm}
\ruimte
An important feature of the above algorithm is that we use $\ell_2=\wdeg g_2$ to decide how many errors to decode. Indeed, it follows from the above lemma that it is not possible to perform list decoding for $t < \ell_2 -k+1$. We now present the main theorem of this section.
\begin{thm}\label{th_main_division}
Let $\mathbf{r} = ( r_1 , \cdots , r_n ) \in \F^n$ be a received word and let $M (\mathbf{r} )$ be the corresponding interpolation module. Let $\{ g_1 , g_2 \}$ be a $(0,k-1 )$-weighted {\tt top} minimal Gr\"obner basis for $M (\mathbf{r} )$ with $\lpos (g_2 ) = 2$. Write $g_2 = \twee{g^{(1)}_2}{g^{(2)}_2}$. Then Algorithm~\ref{alg_main} yields a list of {\em all} message polynomials $m$ such that
\beq
d_H(\mathbf{c}, \mathbf{r}) \label{eq_dist}\;\mbox{is minimal, where $\mathbf{c}=(m(x_1), \ldots, m(x_n))$}.
\eeq
In particular, in case there exists an error pattern with only $\leq \lfloor (n-k)/2 \rfloor$ errors, the list consists of only 
\beq
m= -\frac{g^{(1)}_2}{g^{(2)}_2} . \label{eq_unique}
\eeq
\end{thm}
\pf
Firstly, it follows immediately from Theorem~\ref{thm_req} and Lemma~\ref{lem_para} that any polynomial $m$ that is output by Algorithm~\ref{alg_main} has to have degree $<k$ and satisfy~(\ref{eq_dist}). Vice versa, if $m$ is a polynomial of degree $< k$ that satisfies~(\ref{eq_dist}) then it follows from Lemma~\ref{lem_para} that it must be in the output list of Algorithm~\ref{alg_main}. Finally, let us assume that there are only $\leq \lfloor (n-k)/2 \rfloor$ errors. This implies that there exists a vector $f = \twee {f^{(1)}}{f^{(2)}}$ in $M (\mathbf{r} )$ with $\wdeg f \leq \lfloor (n-k)/2 \rfloor +k-1 < (n+k-1)/2$ that satisfies the requirements of Theorem~\ref{thm_req}. Because of Lemma~\ref{lem_para} it follows that $\ell_2 < (n+k-1)/2$. Now, since $\ell_1 + \ell_2 = n+k -1$ by Theorem~\ref{thm_sum}, this implies that $\ell_1 > \ell_2$. As a result, $a = 0$ in step 3), so that step 4) immediately gives the unique solution for $j=0$ as~(\ref{eq_unique}). 
\eind
\pfend
Our next example illustrates the classical decoding scenario, showing that Algorithm~\ref{alg_main} is an extension of existing classical interpolation-based algorithms as in~\cite{leeS08, gao03}.
\begin{example}
Consider the single-error correcting $(7, 5)$ RS code over $GF(7)$. The message polynomial $m(x) = 2x^2 + x + 3$ is encoded as $\mathbf{c}=(m(0), m(1), \ldots, m(6)) = (3, -1, -1, 3, -3, 2, -3)$. Let the received word be $\mathbf{r} = (3, \mathbf{2}, -1, 3, -3, 2, -3)$. Thus an error occurred at locator position 1. The polynomials $\Lag$ and $\Pi$ are computed as $\Lag (x) = -3x^6-3x^5-3x^4-3x^3-x^2-2x+3$ and $\Pi (x) =x^7-x$. Thus the module $M(\mathbf{r})$ is spanned by the rows of the matrix 
\[
 \left( \begin{array}{cc}
x^7-x & 0 \\
-3x^6-3x^5-3x^4-3x^3-x^2-2x+3 & -1
 \end{array} \right). \]
A minimal Gr\"obner basis $\{ g_1 , g_2 \}$ of $M(\mathbf{r})$ with respect to the $(0,4 )$-weighted {\tt top} monomial order is computed as 
\[
\COL \{ g_1 , g_2 \} = \left( \begin{array}{cc}
-3x^6-3x^5-3x^4-3x^3-x^2-2x+3 & -1 \\
2x^3 - x^2 + 2x - 3 & -x+1
\end{array} \right). \]
Thus, in the terminology of Theorem~\ref{th_main_division} we have $g^{(1)}_2 = 2x^3 - x^2 + 2x - 3$ and $g^{(2)}_2 = -x+1$. Applying Algorithm~\ref{alg_main} we determine that $g^{(1)}_2$ is a multiple of $g^{(2)}_2$ and we recover
\[
m(x) = - \frac{g^{(1)}_2}{g^{(2)}_2} = 2x^2+x+3 .
\]
\end{example}
Let us now move on to an example of decoding beyond the classical error bound.
Our approach is particularly feasible for the case that $b =1$ and $a$ is restricted to a constant, as illustrated in the next example. Note that the example is an instance of ``one-step-ahead'' list decoding~\cite{wu08}.

\ruimte
\begin{example}
\label{exa:one-step-grob}
Consider the single-error correcting $(7, 4)$ RS code over $GF(7)$; let the message polynomial be $m(x) = 2x^2 + x + 3$ which is encoded as $\mathbf{c} = (m(0), m(1), \ldots, m(6)) = (3, -1, -1, 3, -3, 2, -3)$. Let the received word be $\mathbf{r} = (3, \mathbf{2}, -1, 3, \mathbf{2}, 2, -3)$ which differs from $\mathbf{c}$ at locations 1 and 4. The polynomials $\Lag$ and $\Pi$ are computed as $\Lag(x)=-x^6 - 2x^5 + x^4 - x^3 + 2x +3$ and $\Pi(x) = x^7-x$. The interpolation module $M(\mathbf{r})$ is spanned by the rows of the matrix
\[ M(\mathbf{r}) =  \left( \begin{array}{cc}
x^7-x & 0 \\
-x^6 - 2x^5 + x^4 - x^3 + 2x +3 & -1
 \end{array} \right) . \] A minimal Gr\"obner basis $\{ g_1 , g_2 \}$ of $M(\mathbf{r})$ with respect to the $(0,3 )$-weighted {\tt top} monomial ordering is computed as
 \[
 \COL \{ g_1 , g_2 \} =  \left( \begin{array}{cc}

x^5-2x^4-x^3-x^2+x+3&  -3x-1 \\
-2x^4+2x^3+x^2-3x+2 & x^2+2x-3\end{array} \right) . \]
Thus in this example $\ell_1 = \ell_2 =5$, so that $a$ is a constant. Applying Algorithm~\ref{alg_main}, we consider $f = a g_1 + g_2$ for $a=0, \ldots, 6$. Writing $f=\twee{f^{(1)}}{f^2}$, we find that $f^{(2)}$ divides $f^{(1)}$ for $a=0, 2,$ and $4$, giving a list of three message polynomials---we recover not only $m(x)= 2x^2+x+3$ (for $a = 0$), but also the message polynomials $3x^3-2x^2+3x-2$ (for $a = 2$), and $-2x^3-2x^2+3x + 3$ (for $a = 4$).
\end{example}
%%%%%%%%%%%%%%%%%%%%%%%%%%%%%%%%%%%%%%
\subsection{Computation of $g_1$ and $g_2$}
%%%%%%%%%%%%%%%%%%%%%%%%%%%%%%%%%%%%%%
There are various ways in which the required minimal Gr\"obner basis $\{ g_1 , g_2 \}$ of the interpolation module $M(\mathbf{r})$ can be computed. One obvious way is to simply run an existing computer algebra system such as {\sc Singular}, specifying the required $(0,k-1)$-weighted {\tt top} order.
\ruimte
Because of the specific form of $M(\mathbf{r})$ a more efficient way is to apply the Euclidean algorithm to the polynomials $\Pi$ and $\Lag$. More specifically, we have the following algorithm.
\begin{algorithm}[!htb]
%\begin{algorithm}[!htb]
\caption{Computation of $g_1$ and $g_2$ via Euclidean algorithm}
\label{alg_main_euclid}
\begin{algorithmic}
\STATE
\STATE {\bf Input}: Received word $\mathbf{r}= (r_1, \ldots, r_n )$; polynomials $\Pi$ and $\Lag$ given by~(\ref{eq_pi}) and~(\ref{eq_Lagrange}).
\STATE {\bf Output}: Polynomials $g_1$ and $g_2$ in $\F[x]^2$, such that $\{ g_1 , g_2 \}$ is a minimal Gr\"obner basis of $M(\mathbf{r}) $ with respect to the $(0,k-1 )$-weighted {\tt top} monomial order, with $\lpos (g_2) = 2$.
\STATE
\STATE 1. Define polynomials $h_0$, $h_1$, $t_0$ and $t_1$ in $\F[x]$ as
\[
\bmat{cc} h_0 & t_0 \\ h_1 & t_1 \emat := \bmat{cc} \Pi & 0 \\ \Lag & -1 \emat ;
\]
set $j:=0$.
\STATE 2. Check
\beq
\DEG t_{j+1} + k-1 \geq \DEG h_{j+1}; \label{eq_stop}
\eeq
if NO, go to Step 3. If YES, define $g_1 := \twee{h_j}{t_j}$ and $g_2 := \twee{h_{j+1}}{t_{j+1}}$ and STOP.
\STATE 3. Apply the Euclidean algorithm to $h_j$ and $h_{j+1}$, yielding $h_j = q_{j+1}h_{j+1} + h_{j+2}$, where $\DEG h_{j+2} < \DEG h_{j+1}$.
\STATE 4. Write
\[
\bmat{cc} h_{j+1} & t_{j+1} \\ h_{j+2} & t_{j+2} \emat := \bmat{cc} 0 & 1\\ 1& -q_{j+1} \emat \bmat{cc} h_j & t_j \\ h_{j+1} & t_{j+1} \emat ; 
\]
increase $j$ by $1$ and go back to Step 2.
\end{algorithmic}
\end{algorithm}
\begin{thm}\label{th_euclid}
Let $\mathbf{r} = ( r_1 , \cdots , r_n ) \in \F^n$ be a received word and let $M (\mathbf{r} )$ be the corresponding interpolation module. Then Algorithm~\ref{alg_main_euclid} yields a $(0,k-1 )$-weighted {\tt top} minimal Gr\"obner basis $\{ g_1 , g_2 \}$ for $M (\mathbf{r} )$ with $\lpos (g_2 ) = 2$. 
\end{thm}
\pf
Firstly we note that the matrix
\[
\bmat{cc} 0 & 1\\ 1& -q_{j+1} \emat
\]
is unimodular, i.e., has a polynomial inverse. It then follows that, at each step $j$, the rows of the matrix 
\beq
\bmat{cc} h_j & t_j \\ h_{j+1} & t_{j+1} \emat\label{eq_htj}
\eeq
are a {\tt pot} minimal Gr\"obner basis for $M (\mathbf{r} )$ whose $(0,k-1)$-weighted {\tt pot} degrees add up to $n+k-1$. By definition, with respect to the $(0,k-1 )$-weighted {\tt top} order both these row vectors have leading position $1$, until the stopping condition~(\ref{eq_stop}) is met. At this point the second row vector has leading position $2$ and the sum of the $(0,k-1)$-weighted {\tt top} degrees add up to $n+k-1$. It now follows from Corollary~\ref{cor_sum} that the rows of the matrix~(\ref{eq_htj}) must be a $(0,k-1)$-weighted minimal {\tt top} Gr\"obner basis for $M (\mathbf{r} )$.  
\eind\pfend
Yet another alternative is to use an iterative method, interpolating the $x_i$'s step by step for $i=1, \ldots , n$. This method has the advantage that the Lagrange polynomial $\Lag$ does not need to be computed upfront.\begin{algorithm}[!htb]
%\begin{algorithm}[!htb]
\caption{Computation of $g_1$ and $g_2$ via iterative algorithm}
\label{alg_main_iter}
\begin{algorithmic}
\STATE
\STATE {\bf Input}: Received word $\mathbf{r}= (r_1, \ldots, r_n)$.
\STATE {\bf Output}: Polynomials $g_1$ and $g_2$ in $\F[x]^2$, such that $\{ g_1 , g_2 \}$ is a minimal Gr\"obner basis of $M(\mathbf{r}) $ with respect to the $(0,k-1 )$-weighted {\tt top} monomial order, with $\lpos (g_2) = 2$.
\STATE
\STATE 1. Initialize $L_0 := k-1$ and $R_0 := I \in \F^{2\times 2}$; denote $R_j := \bmat{cc} Q_j & -K_j \\ N_j & -D_j \emat\in \F[x]^{2\times 2}$ for $j=0 ,\ldots , n$.
\STATE 2. Process the received values $r_j$ iteratively for $j=1$ to $n$ as follows. For $j=1$ to $n$ do

\begin{enumerate}
\item compute $\Gamma_j := Q_{j-1}(x_j) -r_j K_{j-1}(x_j)$ and $\Delta_j := N_{j-1}(x_j) -r_j D_{j-1}(x_j)$
\item define $R_j := V_j R_{j-1}$, where 
\begin{itemize}
\item $V_j := \bmat{cc} \Delta_j & -\Gamma_j \\ 0 & x-x_j \emat$ and $L_j := L_{j-1}+1$ if $\Delta_j \neq 0$ and ($L_{j-1} < (j+k-1)/2$ or $\Gamma_j = 0$),
\item $V_j := \bmat{cc} x-x_j & 0 \\ \Delta_j & -\Gamma_j \emat$ and $L_j := L_{j-1}$ otherwise
\end{itemize}
\end{enumerate}

\STATE 3. Define $g_1 := \twee{Q_n}{-K_n}$ and $g_2 := \twee{N_n}{-D_n}$.
\end{algorithmic}
\end{algorithm}
\begin{thm}\label{th_iter}
Let $\mathbf{r} = ( r_1 , \cdots , r_n ) \in \F^n$ be a received word and let $M (\mathbf{r} )$ be the corresponding interpolation module. Then Algorithm~\ref{alg_main_iter} yields a $(0,k-1 )$-weighted {\tt top} minimal Gr\"obner basis $\{ g_1 , g_2 \}$ for $M (\mathbf{r} )$ with $\lpos (g_2 ) = 2$. 
\end{thm}
\pf
For $j=1, \ldots , n$ denote the interpolation module associated with $r_1 , \ldots , r_j$ by $M(\rbold_1 , \ldots , \rbold_j )$. We show that the rows of $R_j$ are a Gr\"obner basis of $M(\rbold_1 , \ldots , \rbold_j )$of the required form for $j=1, \ldots , n$. We interpret $L_j$ as the $(0,k-1)$-weighted {\tt top} degree of the second row of $R_j$. Clearly this is true for $j=1$. Let us now proceed by induction and assume that this is true for $j-1 \in \{0,\ldots , n-1\}$. By definition of $V_j$ and the induction assumption the rows of $R_j$ are a Gr\"obner basis for $M(\rbold_1 , \ldots , \rbold_j )$. Also, by construction, their $(0,k-1)$-weighted {\tt top} degrees add up to $1$ more than the $(0,k-1)$-weighted {\tt top} degrees of $R_{j-1}$. Then, by induction, the $(0,k-1)$-weighted {\tt top} degrees of $R_j$ add up to $j+k-1 = \WDEG (M(\rbold_1 , \ldots , \rbold_j ))$. It then follows from Corollary~\ref{cor_sum} that the rows of $R_j$ are a $(0,k-1 )$-weighted {\tt top} minimal Gr\"obner basis for $M(\rbold_1 , \ldots , \rbold_j )$. Finally, by construction and the induction hypothesis, it is easily seen that the second row of $R_j$ has leading position $2$. This proves the theorem.
\eind
%\pfend
%%%%%%%%%%%%%%%%%%%%%%%%%%%%%%%%%%%%%%
\subsection{The special case $\rbold = (y_1 , \ldots , y_{n-k} , 0 , \cdots , 0)$}\label{subsec_reencoding} 
%%%%%%%%%%%%%%%%%%%%%%%%%%%%%%%%%%%%%%
In this subsection we pay special attention to the case that the received word $\rbold$ is of the form $(y_1 , \ldots , y_{n-k} , 0 , \cdots , 0) \in \F^n$. This comes about when so-called "re-encoding" is used in advance of RS decoding, see e.g.,~\cite{kotterVitw03,koetterMV10}.
\ruimte
First we introduce  the polynomial $G \in \F[x]$ of degree $k-1$ as
\beq
G := \prod_{i=n-k+2}^n (x- x_i) .\label{eq_G}
\eeq
Clearly, the polynomials $\Pi$ and $\Lag$ of the previous subsection can be written as
\beq
\Pi = \Pi_y G \label{eq_Piy}
\eeq
and
\beq
\Lag = \Lag_y G ,\label{eq_Lagy}
\eeq
where $\Pi_y$ and $\Lag_y$ are in $\F[x]$.
The following lemma is straightforward.
\begin{lem}
Let $(y_1 , \ldots , y_{n-k})\in \F^{n-k}$, $\rbold = (y_1 , \ldots , y_{n-k} , 0 , \cdots , 0) \in \F^n$ and let $\Pi, \Lag,G,\Pi_y$ and $\Lag_y$ be defined as before. Let $M(\rbold ) := \SPAN \{ \twee{\Pi}{0} , \twee{\Lag}{-1} \}$ as before and define $M^\star (\ybold) := \SPAN \{ \twee{\Pi_y}{0} , \twee{\Lag_y}{-1} \}$. Then the following two statements are equivalent:
\begin{itemize}
\item $\{ g_1 , g_2 \}$ is a minimal Gr\"obner basis of $M^\star(\mathbf{y}) $ with respect to the unweighted {\tt top} order, with $\lpos (g_2) = 2$
\item $\{ \tilde g_1 , \tilde g_2 \}$ is a minimal Gr\"obner basis of $M(\mathbf{r}) $ with respect to the $(0,k-1)$-weighted {\tt top} order, with $\lpos (\tilde g_2) = 2$,
\end{itemize}
where $g_i = \twee{g_i^{(1)}}{g_i^{(2)}}$ and $\tilde g_i = \twee{g_i^{(1)} G}{g_i^{(2)}}$ for $i=1,2$.
\end{lem}
Because of the above lemma it is now straightforward to modify Algorithm~\ref{alg_main} into Algorithm~\ref{Algo-classical-grob}.
\begin{algorithm}[h]
%\begin{algorithm}[!htb]
\caption{Minimal list decoding of $(n, k)$ RS code for re-encoded received word}
\label{Algo-classical-grob}
\label{alg_main_y}
\begin{algorithmic}
\STATE
\STATE {\bf Input}: Received word $\mathbf{y}= (y_1, \ldots, y_{n-k})$ in $\F^{n-k}$.
\STATE {\bf Output}: A list of polynomials $m$ of degree $< k$ such that $d_H(\mathbf{c}, \mathbf{r})$ is minimal, where $\mathbf{c}=(m(x_1), \ldots, m(x_n))$ and $\mathbf{r}= (y_1, \ldots, y_{n-k},0. \ldots ,0)$ in $\F^n$.
\STATE
\STATE 1. Compute the polynomials $\Pi_y$ and $\Lag_y$ given by~(\ref{eq_Piy}) and~(\ref{eq_Lagy}) ; define the interpolation module $M(\mathbf{y}) :=  \SPAN \{ \twee{\Pi_y}{0} , \twee{\Lag_y}{-1}\}$.
\STATE 2. Compute a minimal Gr\"obner basis $G = \{g_1 , g_2\}$ of $M(\mathbf{y}) $ with respect to the unweighted {\tt top} monomial order, with $\lpos (g_2) = 2$; set $j=0$. 
\STATE 3. Compute $f =a g_1 + b g_2$, for all $a \in \F [x]$ with $\deg a \leq \ell_2 - \ell_1 +j$ and for all monic $b \in \F [x]$ with $\deg b = j$; write $f = \twee{f^{(1)}}{f^{(2)}}$. Check whether $f^{(1)} G$ is a multiple of $f^{(2)}$, where $G$ is given by~(\ref{eq_G}).
\STATE 4. Whenever step 3) is successful, output all obtained quotient polynomials, i.e., polynomials $m$ of the form $m= -f^{(1)} G / f^{(2)}$. In case step 3) is not successful increase $j$ by $1$ and repeat step 3).
\end{algorithmic}
\end{algorithm}
\ruimte

Again the Euclidean algorithm can be used to compute $g_1$ and $g_2$; for this, Algorithm~\ref{alg_main_euclid} should be initialized by $\Pi_y$ and $\Lag_y$ instead of $\Pi$ and $\Lag$ and the stopping criterion~(\ref{eq_stop}) should be replaced by
\[
\DEG t_{j+1}\geq \DEG h_{j+1},
\]
instead of~(\ref{eq_stop}). 
\ruimte
An alternative way to compute $g_1$ and $g_2$ is to employ an algorithm that processes the values of $y_1, \ldots , y_{n-k}$ iteratively. For this, Algorithm~\ref{alg_main_iter} is modified into Algorithm~\ref{alg_main_iter_re-encoding} which essentially coincides with the well-known Welch-Berlekamp algorithm~\cite{welchWB86}, see also~\cite{kuijisit00,kuijima01}.
\begin{algorithm}[!htb]
%\begin{algorithm}[!htb]
\caption{Computation of $g_1$ and $g_2$ via iterative algorithm for re-encoded received word}
\label{alg_main_iter_re-encoding}
\begin{algorithmic}
\STATE
\STATE {\bf Input}: Received word $\mathbf{y}= (y_1, \ldots, y_{n-k})$ in $\F^{n-k}$.
\STATE {\bf Output}: Polynomials $g_1$ and $g_2$ in $\F[x]^2$, such that $\{ g_1 , g_2 \}$ is a minimal Gr\"obner basis of $M(\mathbf{y}) $ with respect to the unweighted {\tt top} monomial order, with $\lpos (g_2) = 2$.
\STATE
\STATE 1. Denote $R_j := \bmat{cc} Q_j & -K_j \\ N_j & -D_j \emat$ for $j=0 ,\ldots , n$; initialize $L_0 := 0$ and 
\[
R_0 := \bmat{cc} x-x_{n-k+1} & 0 \\ 0 & 1 \emat \in \F[x]^{2\times 2}
\]
\STATE 2. Process the received values $y_j$ iteratively for $j=1$ to $n-k$ as follows. For $j=1$ to $n-k$ do
\begin{enumerate}
\item compute $\Gamma_j := Q_{j-1}(x_j) -r_j K_{j-1}(x_j)$ and $\Delta_j := N_{j-1}(x_j) -r_j D_{j-1}(x_j)$
\item define $R_j := V_j R_{j-1}$, where 
\begin{itemize}
\item $V_j := \bmat{cc} \Delta_j & -\Gamma_j \\ 0 & x-x_j \emat$ and $L_j := L_{j-1}+1$ if $\Delta_j \neq 0$ and ($L_{j-1} < j/2$ or $\Gamma_j = 0$),
\item $V_j := \bmat{cc} x-x_j & 0 \\ \Delta_j & -\Gamma_j \emat$ and $L_j := L_{j-1}$ otherwise
\end{itemize}
\end{enumerate}
\STATE 3. Define $g_1 := \twee{Q_{n-k}}{-K_{n-k}}$ and $g_2 := \twee{N_{n-k}}{-D_{n-k}}$.
\end{algorithmic}
\end{algorithm}
\ruimte
%%%%%%%%%%%%%%%%%%%%%%%%%%%%%%%%%%%%%%
\section{Minimal list decoding through rational interpolation}
\label{sec:rational}
%%%%%%%%%%%%%%%%%%%%%%%%%%%%%%%%%%%%%%
The most computationally intensive task in Algorithm~\ref{alg_main} is Step 3. Recall that in Step 3, we need to determine all $a$ and $b$ of degree $k_1 \leq \ell_2-\ell_1+j$ and $k_2=j$ such that $f^{(1)}$ is a multiple of $f^{(2)}$. A brute force approach may be to consider
\[
f=\twee{f^{(1)}}{f^{(2)}} = a \twee{g_1^{(1)}}{g_1^{(2)}} + b \twee{g_2^{(1)}}{g_2^{(2)}}
\]
and check for all polynomials $a$ and $b$ of bounded degree $k_1$ and $k_2$, respectively, whether $f^{(2)}$ divides $f^{(1)}$. Clearly this approach is feasible only when both $k_1$ and $k_2$ are small. For large values of $k_1$ and $k_2$, the computational complexity becomes prohibitively high, especially when the code is defined over a large field. Fortunately, Step 3 can be formulated as an algebraic curve fitting problem for which efficient polynomial time algorithms exist. We explain this approach in the following.

It follows from Theorem~\ref{thm_req} that, in the context of Algorithm~\ref{alg_main}, $f^{(1)}$ is a multiple of $f^{(2)}$ if and only if $f^{(2)}$ has 
$t=\ell_2 -k + 1 + j$ distinct roots. Therefore, an alternative approach to Step 3 is to determine all $a$ and $b$ of degree $k_1 \leq t + k - \ell_1 - 1$ and $k_2 = t + k - \ell_2-1$, respectively, such that 
\begin{equation}
\label{eqn:error-loc}
f^{(2)}(x) = a(x)g_1^{(2)}(x) + b(x)g_2^{(2)}(x)
\end{equation}
has $t$ distinct roots. Now dividing both sides of~(\ref{eqn:error-loc}) by $g_1^{(2)}(x)$ we get
\begin{equation}
\frac{f^{(2)}(x)}{g_1^{(2)}(x)} = a(x) + b(x) \frac{g_2^{(2)}(x)}{g_1^{(2)}(x)}.
\end{equation}
Now let us define
\[
 z_i = - \frac{g_2^{(2)}(x_i)}{g_1^{(2)}(x_i)}, \quad \textnormal{for $i=1, \cdots, n$}.
\]
Then Step 3 of Algorithm~\ref{alg_main} can be formulated as the following rational interpolation problem.

\ruimte
{\bf Rational Interpolation Problem}: Given $n$ points $(x_1, z_1), (x_2, z_2) \cdots, (x_n, z_n)$ and a non-negative integer $t$, determine all rational polynomials of the form $z = a/b$, with $a$ and $b$ of degree $k_1$ and $k_2$, respectively, such that $z$ passes through $t$ of the $n$ points $(x_1, z_1), (x_2, z_2) \cdots, (x_n, z_n)$. 
\ruimte
This problem looks similar to the interpolation problem addressed by Guruswami and Sudan in~\cite{gursud99}. However, it is complicated by the fact that now we look for a rational solution rather than a polynomial solution. Recently, this rational interpolation problem has been addressed by Wu in~\cite{wu08}. For the sake of completeness we briefly describe Wu's formulation here.

\subsection{Wu's rational interpolation algorithm}
In line with the Guruswami-Sudan approach, Wu's algorithm first computes a bivariate polynomial $Q(x,z)$, satisfying certain constraints, that passes through all the $n$ points $(x_1, z_1), (x_2, z_2) \cdots, (x_n, z_n)$. Then the desired rational solutions $z=a/b$ are obtained from the factorization of $Q(x, z)$. Given the values of $t$, $k_1$, and $k_2$,  let us determine the constraints that must be satisfied for the existence of such a $Q(x, z)$. 

Let us define the $(1, w)$ weighted degree of a bivariate polynomial $Q(x, z) = \sum_{(i,j) \in I}a_{i,j}x^iz^j$ as
\begin{equation}
\wdeg_{1,w}Q(x,z) = \max_{(i,j) \in I} \{i+jw\}.
\end{equation}
Let $w:=k_1 - k_2$, $\rho := \wdeg_{(1,w)}Q(x,z)$, and  $M:=\wdeg_{0, 1}Q(x,z)$. Clearly $\deg_{0, 1}Q(x,z)$ is the $z$-degree of $Q(x,z)$. Now if $z=a/b$ passes through $t$ points with multiplicity $s$ then the polynomial $b(x)^{M}Q(x, a(x)/b(x))$ must have $ts$ roots. On the other hand, $b(x)z - a(x)$ will be a factor of $Q(x,z)$ if $b(x)^M Q(x, a(x)/b(x))$ is identically zero. In turn, $b(x)^{M}Q(x,a(x)/b(x))$ will be identically zero if it has more roots than its degree. Now the degree of $b(x)^{M}Q(x, a(x)/b(x))$ is at most $\rho + M k_2$ Therefore, a necessary condition that must be satisfied is
\begin{equation}
\label{eqn:degree-constraint}
\rho + M k_2 < ts.
\end{equation}

On the other hand, a necessary condition for the existence of $Q(x,z)$ passing through the $n$ points with multiplicity $s$ is that its $(u,v)$-th Hasse derivatives at all the $n$ points are zero for all $u+v \leq s$. Thus the requirement that $(x_i, z_i)$ be a zero of $Q(x,z)$ with multiplicity $s$, for all $i=1, 2, \ldots, n$, leads to $N$ constraints in the form of $N$ homogeneous equations where
\begin{equation}
\label{eqn:numconst}
 N = ns(s+1)/2
\end{equation}
and unknown variables are the coefficients of $Q(x,z)$.  A nonzero solution to the system of homogeneous equations is guaranteed to exist if the number of equations is less than the number of unknowns. Now the number of coefficients in $Q(x,z)$ with $\wdeg_{1,w} Q(x,z) = \rho$ and $\wdeg_{0,1}Q(x,z)=M$ is
\begin{equation}
\label{eqn:numunknown}
U = (\rho+1)(M+1) - \frac{w}{2} M (M+1).
\end{equation}
Therefore, a sufficient condition for the existence of a $Q(x,z)$, passing through all the $n$ points with multiplicity $s$, is 
\begin{equation}
\label{eqn:unknown-constraint}
(\rho+1)(M+1) - \frac{w}{2} M (M+1) > \frac{ns(s+1)}{2}.
\end{equation}
 
Wu, in~\cite{wu08}, has proposed suitable choices for the values of $s$, $M$, and $\rho$ satisfying~(\ref{eqn:degree-constraint}) and~(\ref{eqn:unknown-constraint}) as
\begin{eqnarray}
\label{eqn:multiplicity}
s &=& \left \lfloor \frac{t(n-k+1-t)}{t^2 - n(2t-(n-k+1))} \right \rfloor, \\
\label{eqn:list_size}
M &=& \left \lfloor \frac{st}{2t-(n-k+1)} \right \rfloor, \\
\label{eqn:wdegree}
\rho & = & ts - M k_2 - 1.
\end{eqnarray}
%With these values of $s$ and $M$, Wu's algorithm attempts to construct the bivariate polynomial $Q(x,z)$ with $\wdeg_{1,w}Q(x,z) = ts - M k_2 -1$. 
For more details on Wu's algorithm see~\cite{JohanMastersThesis}. It is worth noting that the multiplicity $s$, computed using~(\ref{eqn:multiplicity}), is not minimal. Although Wu suggested to first compute $s$ according to~(\ref{eqn:multiplicity}) and then greedily minimize it subject to a certain constraint, he did not give any explicit algorithm to compute the minimal value of $s$. More importantly, given the minimum $s$, the values of $M$ and $\rho$ computed in~(\ref{eqn:list_size}) and~(\ref{eqn:wdegree}) are not necessarily optimal. In the next section, we present an algorithm that computes the minimum value of $s$ as well as the associated optimal values of $M$ and $\rho$.

\subsection{Optimizing the integer parameters}
Given feasible values of $s$, $M$, and $\rho$, the rational interpolation step involves two steps: (1) construction of $Q(x,z)$ and (2) factorization of $Q(x,z)$. The best known algorithm for the construction of the interpolating polynomial $Q(x,z)$ is the K\"otter algorithm~\cite{koetterIT96}. The K\"otter algorithm has a complexity of $O(MN^2)$~\cite{koetterMV10}, where $N$ is the number of constraints as defined in~(\ref{eqn:numconst}). More precisely, it has memory complexity of $O(MU)$ and time complexity of $O(NMU)$~\cite{gross02}, where $U$ is the number of coefficients in $Q(x,z)$ as defined in~(\ref{eqn:numunknown}) and $M$ is the $z$-degree of the interpolating polynomial $Q(x,z)$. On the other hand, the rational factorization step can be done in time $O(n^{3/2}s^{7/2})$ using Wu's rational factorization procedure~\cite{wu08}. As analyzed in sub-section~\ref{subsec:complexity}, it is the K\"otter algorithm that dominates the overall memory and computational complexity of the proposed, as well as Wu's, list decoding procedure. Therefore, to reduce the complexity of the K\"otter algorithm, we take the following two step strategy. In the first step, we derive an explicit method to determine the minimum value of $s$ for which there exist some $M$ and $\rho$ satisfying~(\ref{eqn:degree-constraint}) and~(\ref{eqn:unknown-constraint}). Once the minimum multiplicity is determined, $N$ becomes fixed. Then in the second step, we compute the optimal values of $M$ and $\rho$ such that $MU$ is minimized.
\ruimte
The constraint~(\ref{eqn:degree-constraint}) can be geometrically interpreted as follows. Assume that $t$ and $s$ are fixed. With the requirement that all the values involved in~(\ref{eqn:degree-constraint}) are non-negative integers, all feasible values of $\rho$ and $M$ must be on or below the line $L$ defined by the equation
\begin{equation}
\label{eqn:equality-multi-minus}
\rho + M k_2 = ts-1.
\end{equation}

On the other hand, the constraint~(\ref{eqn:unknown-constraint}) requires that all feasible values of $\rho$ and $M$ are above the curve $C$ defined by the equation
\begin{equation}
\label{eqn:equality-const}
(\rho+1)(M+1) - \frac{w}{2} M (M+1) = \frac{ns(s+1)}{2}.
\end{equation}

Therefore, a necessary condition for the existence of a feasible solution satisfying both the constraints~(\ref{eqn:degree-constraint}) and~(\ref{eqn:unknown-constraint}) is that $L$ intersects $C$ at two different points $(M_1, \rho_1)$ and $(M_2, \rho_2)$ on the real plane. Now solving~(\ref{eqn:equality-multi-minus}) and~(\ref{eqn:equality-const}) for $M$ we get
\begin{equation}
\label{eqn:ellasfunck1k2}
M = \frac{(ts-k_0) \pm \sqrt{(ts-k_0)^2 - 4 (N-ts)k_0}}{2k_0},
\end{equation}
where $k_0 = (k_1+k_2)/2$. According to Algorithm~\ref{alg_main}, while correcting $t = \ell_1 - k + 1 + j$ errors, we have $k_1 = \ell_2 - \ell_1 + j$ and $k_2 = j$. Using $\ell_1 + \ell_2 = n+k-1$, we get $k_0 = (t-t_0)$ where $t_0 = d/2$. Substituting $k_0 = (t-t_0)$ in~(\ref{eqn:ellasfunck1k2}) we get
\begin{equation}
\label{eqn:ell-func-ntk-only}
M = \frac{(ts- t + t_0) \pm \sqrt{(ts - t + t_0)^2 - 4 (N-ts)(t-t_0)}}{2(t-t_0)}.
\end{equation}
It follows from~(\ref{eqn:ell-func-ntk-only}) that the value of $M$ and thus the choice of $s$ is independent of $k_1$ and $k_2$.
%\textnormal{where} \; k_0 = \frac{k_1+k_2}{2} \; \textnormal{and} \; {N = \frac{ns(s+1)}{2}}
Now for a fixed $s$, it can easily be verified if $L$ and $C$ intersect at two different points on the real plane by checking whether
\begin{equation}
\label{eqn:positivity}
(ts - t + t_0)^2 > 4 (N-ts)(t-t_0).
\end{equation}
According to~(\ref{eqn:positivity}) any feasible $s$ must satisfy the following inequality which was also derived in Wu~\cite{wu08}
\begin{equation}
\label{eqn:lower-multi}
s^2(t^2-2 (t-t_0) n) - 2s(n-t)(t-t_0) + (t-t_0)^2 > 0.
\end{equation}
This in turn implies that
\begin{equation}
\label{eqn:gen-multi}
s > \frac {(t-t_0)(n-t+\sqrt{n(n-d)})}{t^2-2n(t-t_0)}.
\end{equation}
From~(\ref{eqn:gen-multi}) it also follows that a feasible value of $s$ will exists only if
\begin{equation}
t^2-2n(t-t_0) > 0,
\end{equation}
which also leads to the same bound on the list decoding radius as derived in~\cite{gursud99}
\begin{equation}
\label{eqn:tmax}
t < n - \sqrt{n(n-d)}).
\end{equation}
Also from~({\ref{eqn:gen-multi}) we get the lower bound on $s$ as
\begin{equation}
s_l = \left \lfloor \frac {(t-t_0)(n-t+\sqrt{n(n-d)})}{t^2-2n(t-t_0)} \right \rfloor  + 1.
\end{equation}
Moreover, an upper bound on $s$ was derived in~\cite{wu08} as
\begin{equation}
\label{eqn:upper-s}
s_u = \left \lfloor \frac{t(2t_0-t)}{t^2-2n(t-t_0)} \right \rfloor + 1.
\end{equation}
\begin{figure}%
\centering
\subfloat[][]{
%\caption{(a)}
\includegraphics[width=7cm]{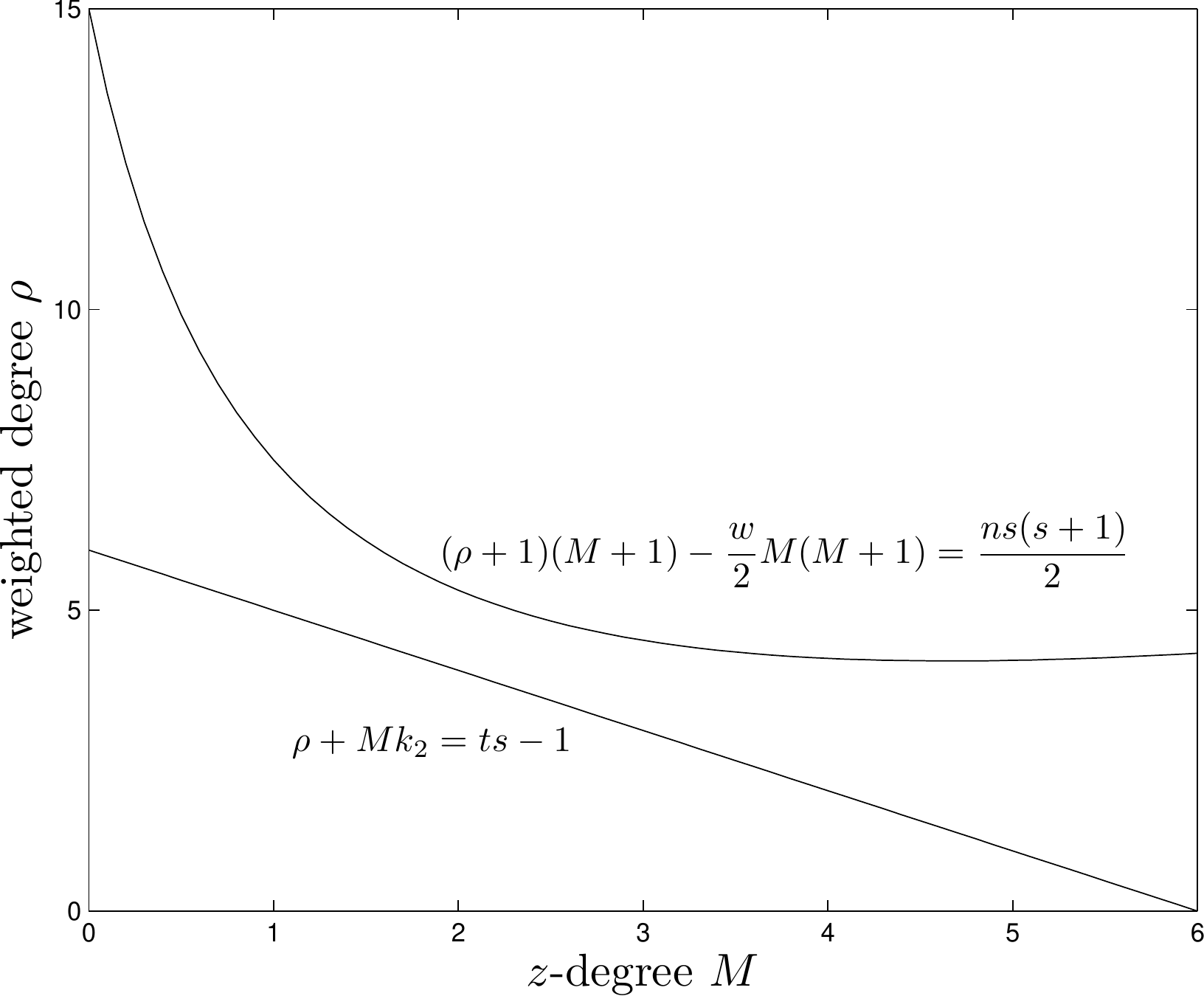}
}%
\qquad
\subfloat[][]{
%\caption{(b)}
\includegraphics[width=7cm]{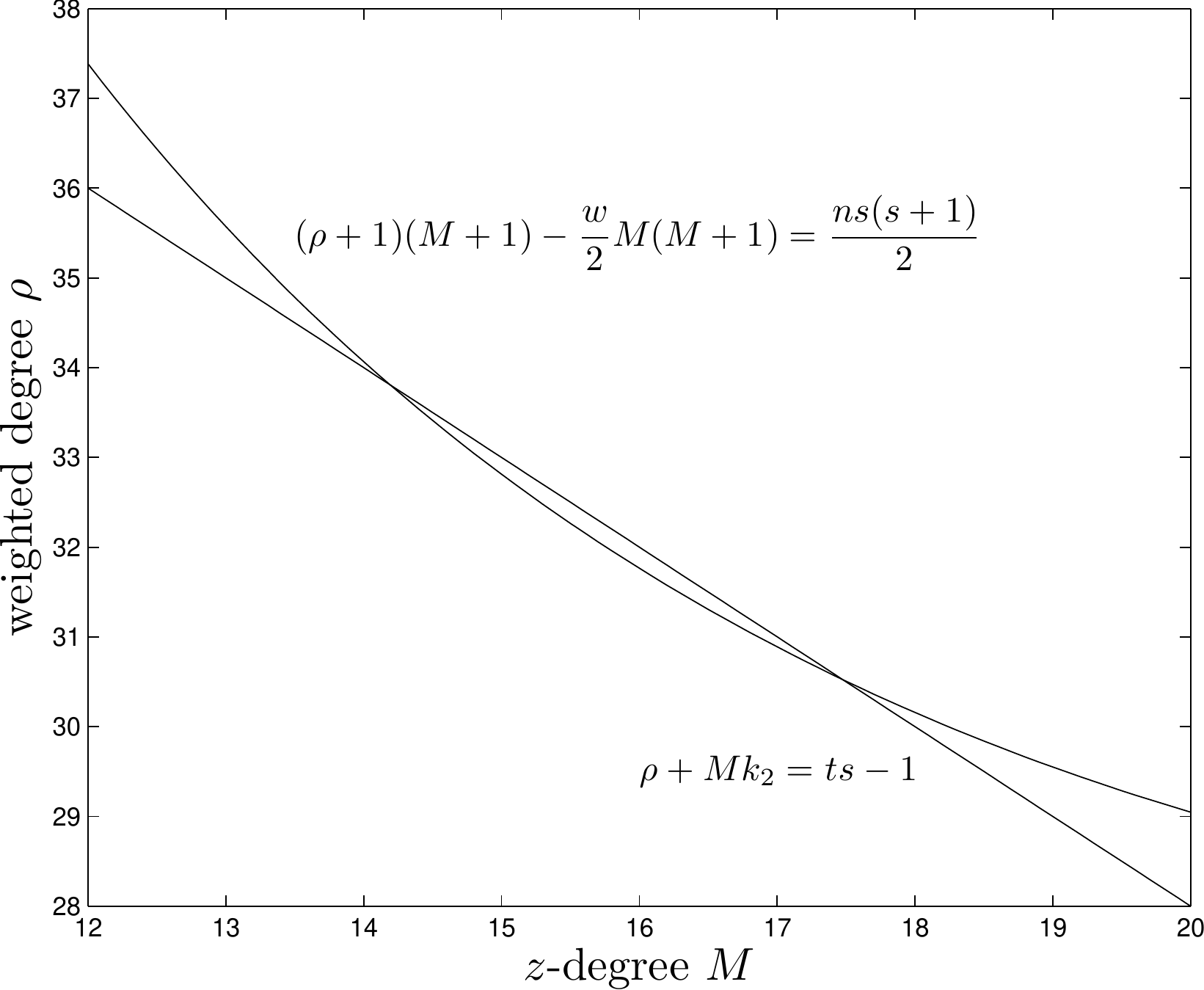}
}
\caption{ Consider correcting $t=7$ errors in the decoding of a $(15, 5)$ RS code over $GF(16)$ when $k_1=2$ and $k_2=1$; (a) With $s=1$, as the line $L$ does not intersect the curve $C$, no feasible values for $M$ and $\rho$ exist; (b) With $s=7$, the line $L$ intersects the curve $C$ at two different points $(M_1=14, \rho_1)$ and $(M_2=17.67, \rho_2)$. Thus $M^*=15, 16, 17$ are feasible choices for $M$. The minimum value of $\rho^*$ corresponding to $M^* = 15$ can be computed as $\rho^* = 33$ since the line $M=15$ intersects $L$ and $C$ at $(15, \rho_h = 33)$ and $(15, \rho_l = 32.81)$, respectively.}
\label{fig:fes-or-not-fes}%
\end{figure}
%Similarly for $M=16$ and $M=17$ we have $\rho^*=32$ and $\rho^*=31$ respectively. Since for $s<7$ no feasible values for $M$ and $\rho$ exists, we get $s_{\min} = 7$, $M_{\textnormal{opt}}=15$ and $\rho_{\textnormal{opt}}=33$.

Thus any $s$, such that $s_l \leq s \leq s_u$ will satisfy the condition~(\ref{eqn:positivity}). Now assume that for a particular $s$, the condition~(\ref{eqn:positivity}) is satisfied, i.e., $L$ and $C$ intersect at two different points $(M_1, \rho_1)$ and $(M_2, \rho_2)$ on the real plane. Without any loss of generality let us assume that $M_1 < M_2$. When $L$ and $C$ intersect at two different points on the real plane, there will exist a feasible solution if there is an integer $M^*$ such that $M_1 < M^* < M_2$, i.e., if 
\begin{equation}
\label{eqn:minell}
\lfloor M_1 \rfloor + 1< M_2.
\end{equation}
Clearly if~(\ref{eqn:minell}) is satisfied, then any $M^* \in \left [\lfloor M_1 \rfloor + 1, \lceil M_2 \rceil -1 \right ]$ is a feasible choice of $M$. Now according to~(\ref{eqn:numunknown}), for a feasible choice of $M=M^*$, it is desirable to find the minimum value of $\rho$ so that $U$ is minimized. Let the line $M = M^*$ intersect $C$ and $L$ at points $(M^*, \rho_l)$ and $(M^*, \rho_h)$ respectively. Since $L$ intersects $C$ from above, it must be the case that $\rho_l < \rho_h$. Although $\rho_h$ is a feasible choice for $\rho$, as used by Wu, we choose the minimum possible value as
\begin{equation}
\rho^* = \lfloor \rho_l \rfloor + 1.
\end{equation}
We illustrate the method of computing the feasible values of the integer parameters, using a particular example, in Fig.~\ref{fig:fes-or-not-fes}.
\ruimte

Now to find the optimal value of $M$ and $\rho$ such that $MU$ is minimized, we need to compute $\rho^*$ and $U^*$ for all $M^* \in \left [ \lfloor M_1 \rfloor + 1, \lceil M_2 \rceil - 1 \right ]$
and choose $M^*$ and $\rho^*$ that result in the minimum value of $MU$. We summarize the above procedure in Algorithm~\ref{alg:min-params} that computes the values of minimum multiplicity $s_{\min}$ and the associated optimal $z$-degree $M_{\textnormal{opt}}$ and weighted degree $\rho_{\textnormal{opt}}$.
\begin{algorithm}[!bht]
%\begin{algorithm}[!htb]
\caption{Compute optimal values of the integer parameters}
\label{alg:min-params}
\begin{algorithmic}
\STATE
\STATE {\bf Input}: $n$, $k$, $t$, $k_1$, and $k_2$.
\STATE {\bf Output}: Minimum multiplicity $s_{\min}$ and optimal $z$-degree $M_{\textnormal{opt}}$ and weighted degree $\rho_{\textnormal{opt}}$.
\STATE
\STATE Compute $w := k_1-k_2$, $d := n-k+1$, $t_0 := d/2$.
\STATE Initialize $s := \max( s_l = \lfloor {(t-t_0)(n-t+\sqrt{n(n-d)})}/{(t^2-2n(t-t_0))} \rfloor  + 1,1)$
\STATE \qquad \qquad $M_{\textnormal{opt}}:=\infty$, $\rho_{\textnormal{opt}}:=\infty$, $U_{\textnormal{opt}}=\infty$
\WHILE{no feasible solution is found}
\STATE Compute $N := ns(s+1)/2$.
\IF{$(ts-t+t_0)^2 > 4(N-ts)(t-t_0)$}
\STATE $(M_2, M_1) := {((ts- t + t_0) \pm \sqrt{(ts - t + t_0)^2 - 4 (N-ts)(t-t_0)})}/{2(t-t_0)}$
\IF{$\lfloor M_1 \rfloor + 1 < M_2$}
\STATE $s_{\min} := s$
\FOR{$M=\lfloor M_1 \rfloor + 1$ to $\lceil M_2 \rceil -1$}
\STATE $\rho := \lfloor N/(M+1) + w/2 M - 1 \rfloor + 1$
\STATE $U := (\rho+1)(M+1) - w/2 M (M+1)$
\IF{$MU < M_{\textnormal{opt}}U_{\textnormal{opt}}$}
\STATE $M_{\textnormal{opt}}:=M$, $\rho_{\textnormal{opt}}:=\rho$, $U_{\textnormal{opt}}=U$
\ENDIF
\ENDFOR
\STATE return $s_{\min}, M_{\textnormal{opt}}, \rho_{\textnormal{opt}}$
\ENDIF
\ENDIF
\STATE $s := s+1$
\ENDWHILE
\end{algorithmic}
\end{algorithm}
\ruimte

{\em Complexity of Algorithm~\ref{alg:min-params}}: The complexity of the algorithm is dominated by the {\bf while} loop and the {\bf for} loop. Number of times the {\bf while} loop is executed is bounded by $s_{\min}$. The {\bf for} loop executes 
\begin{equation}
\label{eqn:boundfor}
O(\sqrt{(ts - t + t_0)^2 - 4 (N-ts)(t-t_0)}/(t-t_0)) =  O(ts)
\end{equation}
times. Moreover, the maximum list decoding radius is $t = \lceil n - \sqrt{n(n-d)}) - 1 \rceil = O(n)$. Thus Algorithm~\ref{alg:min-params} computes the integer parameter values in time $O(ns^2)$.

%After the computation of $s_{\min}, M_{\min}, \rho_{\min}$, the next step is to compute the bivariate polynomial $Q(x, z)$ of $\wdeg_{1,w}Q(x,z) = \rho_{\min}$ and $\wdeg_{0,1}Q(x,z) = M_{\min}$ that passes through the points $(x_i, z_i), i=1, \cdots, n$, with multiplicity $s_{\min}$. The best known algorithm for this task is the K\"otter algorithm~\cite{koetterIT96}.

\subsection{Computation of the message polynomial}
After constructing the bivariate polynomial, the solutions to the rational interpolation problem can be obtained by the rational factorization procedure of~\cite{wu08}. Clearly every solution $(a, b)$ to the rational interpolation problem gives a valid error locator polynomial $f^{(2)}= a g_1^{(2)} + b g_2^{(2)}$. Given a valid error locator polynomial $f^{(2)}$, Wu's algorithm uses Forney's formula to compute the error magnitudes and hence the codeword. However, in our approach, the message polynomial can be computed in a simpler way: for every solution $(a, b)$, it can be computed as \[
m(x) = - \frac{a g_1^{(1)} + b g_2^{(1)}}{a g_1^{(2)} + b g_2^{(2)}}.
\]

\subsection{Complexity}
\label{subsec:complexity}
We summarize the complete minimal list decoding algorithm in Algorithm~\ref{alg:complete}.
\begin{algorithm}[!htb]
%\begin{algorithm}[!htb]
\caption{Minimal list decoding of $(n, k)$ RS code using rational interpolation}
\label{alg:complete}
\begin{algorithmic}
\STATE
\STATE {\bf Input}: Received word $\mathbf{r}= (r_1, \ldots, r_n)$.
\STATE {\bf Output}: A list of polynomials $m$ of degree $< k$ such that $d_H(\mathbf{c}, \mathbf{r})$ is minimal, where $\mathbf{c}=(m(x_1), \ldots, m(x_n))$.
\STATE
%\STATE 1. Compute the polynomials $\Pi$ and $\Lag$ given by~(\ref{eq_pi}) and~(\ref{eq_Lagrange}); Define the interpolation module $M(\mathbf{r}) :=  \SPAN \{ \twee{\Pi}{0} , \twee{\Lag}{-1}\}$.
\STATE 1. Compute a minimal Gr\"obner basis $G = \{g_1 , g_2\}$ of $M(\mathbf{r}) $ with respect to the $(0,k-1 )$-weighted {\tt top} monomial order, with $\lpos (g_2) = 2$ using Algorithm~\ref{alg_main_iter} (or using Algorithm~\ref{alg_main_iter_re-encoding} if re-encoding is used). Denote $\ell_1 := \wdeg g_1$ and $\ell_2 := \wdeg g_2$; set $j=0$.
\STATE 2. With $t := \ell_2 - k + 1 + j$, $k_1 := \ell_2 - \ell_1 +j$, and $k_2 := j$ compute $s_{\min}$, $M_{\textnormal{opt}}$, and $\rho_{\textnormal{opt}}$ using Algorithm~\ref{alg:min-params}.
\STATE 3. Construct $Q(x,z)$ of $\wdeg_{0,1}Q(x,z)=M_{\textnormal{opt}}$ and $\wdeg_{1,w}Q(x,z)=\rho_{\textnormal{opt}}$ passing through $(x_i, z_i)_{i=1}^{n}$, with multiplicity $s_{\min}$ using the K\"otter algorithm from~\cite{McEliece03}.
\STATE 4. Compute all factors of $Q(x,z)$ of the form $z - a/b$ using the rational interpolation algorithm from~\cite{wu08}.
\STATE 5. If step 4 is successful, output all obtained quotient polynomials, i.e., polynomials $m$ of the form $m= -f^{(1)} / f^{(2)}$; Otherwise increase $j$ by $1$ and go to step 3.
\end{algorithmic}
\end{algorithm}
%In step~1, computation of $\Pi$ and $\Lag$ involve $O(n)$ and $O(n\log^2 n\log \log n)$~\cite{beelen10} finite field operations, respectively. 
The computation of the minimal Gr\"obner basis in step 1 using Algorithm~\ref{alg_main_iter} takes $O(n^2)$ operations. Algorithm~\ref{alg:min-params} in step 2 takes $O(ns^2)$ time. The K\"otter algorithm used in step 3 involves $O(MN^2)=O(Mn^2s^4)$ operations~\cite{koetterMV10}, where $N$ is the number of constraints as defined in~(\ref{eqn:numconst}) and $M$ is the $z$-degree of the interpolating polynomial $Q(x,z)$. The rational factorization in step 4 can be done in time $O(n^{3/2}s^{7/2})$~\cite{wu08}. Thus the overall complexity of the proposed algorithm is $O(MN^2)$. However, because of step 2, our list decoding algorithm optimizes $MU$. Since, more precisely, the K\"otter algorithm involves memory complexity of $O(MU)$ and time complexity of $O(NMU)$, our algorithm uses less memory as well as computation as compared to Wu's method. The advantage of the proposed algorithm in terms of $z$-degree $M$ and number of unknown coefficients $U$ is illustrated in Example~\ref{ex:comparision}.
\begin{example}
\label{ex:comparision}
Consider the $(127, 24)$ RS code defined over $GF(2^7)$ with $d = 104$. Consider correcting $t=64$ errors when $k_1=15$ and $k_2=9$. For this instance, Wu's algorithm using~(\ref{eqn:multiplicity}) computes $s=2$, which is also the minimum multiplicity. Now Wu's algorithm computes $M = 5$ and $\rho = 72$ using~(\ref{eqn:list_size}) and~(\ref{eqn:wdegree}), respectively. With these values, Wu's algorithm requires solving a system of $N = 381$ homogeneous equations involving $U = 408$ unknowns. In contrast, in our algorithm we find that when $s_{\min}=2$, the line $L$ intersect the curve $C$ at points $(3.3241,*)$ and $(6.3426, *)$. Now for the feasible values of $M^*=4, 5, 6$, we get $\rho^*=88, 78, 72$ and $U^* = 385, 384, 385$, respectively. Finally we get the optimal values as $M_{\textnormal{opt}}=4$ and $\rho_{\textnormal{opt}}=88$ with $U_{\textnormal{opt}}=385$.
\end{example}

\section{Conclusions}
\label{sec:con}
In this paper we have taken a parametric approach to the problem of minimal list decoding. The proposed algorithms have error correcting radius $L$, where $L$ is the minimum of the Hamming distances between the received word and any codeword in $C$. There are several important features of the approach. Firstly, the minimality of $L$ ensures that all solutions correspond to valid codewords and therefore we do not need to check for validity. The parameterization can also be used for general list decoding, however, then a check on the validity of the corresponding codewords needs to be carried out. Secondly, upon computation of a solution of the rational interpolation problem or, equivalently, of an error locator polynomial, we do not need to determine the error magnitudes via Forney's formula. Instead, solutions to the rational interpolation problem directly lead to message polynomials. Thirdly, we provide a geometric approach to optimize the integer parameters associated with the problem of rational interpolation. Since the interpolation step is the most computationally intensive task in list decoding, optimization of the integer parameters results in less computational as well as memory requirements. Finally, by using re-encoding as in sub-section~\ref{subsec_reencoding}, the approach lends itself well to the type of distributed source coding (DSC) proposed in~\cite{aliK10}.

\section*{Acknowledgment}
We thank Nikeeth Venkatraman Ramanathan for helping in implementing the computational examples.

%\begin{itemize}
%\item list decoding for improved DSC--compression is achieved by re-encoding (essentially evaluating a Lagrange polynomial of degree $< k$), rather than by computing syndrome; is actually erasures-only decoding.
%\item parametrization can also be used for non-minimal list decoding but then a check on the validity of $m$ (or, equivalently, a check on the validity of the roots of $f^2$) needs to be carried out.
%
%\end{itemize}

%%%%%%%%%%%%%%%%%%%%%%%%%%%%%%%%%%%%%%%%%%%%%%%%%%%%%%%%%%%%%%%%%%%%%%%%%%%%%%%%
%%%%%%%%%%%%%% %%%%%%%%%%%%%%%%%%%%
\bibliographystyle{plain}
\bibliography{MA-MK}
%\begin{thebibliography}{99}
%\end{thebibliography}

\end{document}